\documentclass[a4paper, 12pt, reqno]{amsart}
\usepackage{amsmath, amssymb, latexsym, graphicx, hyperref, times}

\newtheorem{proposition}{Proposition}

\newtheorem{definition}{Definition}

\begin{document}

\title{Living being as informed systems:\\ Towards a physical
theory of information}

\maketitle

\begin{center}
\vspace{-24pt}(Journal of Biological Systems (1996), 4, 565-584)

(Revised version: 2006)

\vspace{20pt} Antonio Le\'on

Instituto Francisco Salinas, Salamanca, Spain

E-mail: aleonsanchez@terra.es

\end{center}

\pagestyle{myheadings}

\markboth{Living beings as informed systems: towards a physical
theory of information}{Living beings as informed systems: towards
a physical theory of information}

\begin{abstract}
I propose here a new concept of information based on two relevant
aspects of its expression. The first related to the undeniable
fact that the expression of information modifies the physical
state of its receiver. The second to the arbitrariness of such
physical changes. In fact, these changes are not deducible from
physical laws but from a code established arbitrarily. Thus,
physical information is proposed here as the capacity of producing
arbitrary changes. Once defined physical information from this
physical point of view, I deduce some basic properties of informed
systems. These properties (renewal, self-reproducing, evolution,
diversification) are immediately recognized as the attributes most
characteristic of living beings, the only natural informed systems
we know. I also propose here a double evaluation of information.
The former is an absolute measure of the physical effects of its
expression based on Einstein's probability. The second is a
functional measure based on the probability that an informed
systems attain a given objective as consequence of the expression
of its information.
\end{abstract}

\section{Introduction}
\noindent In scientific literature, especially in Biology, the
word information is used with different meanings
\cite{Hacken1993}, although only one of them, the probabilistic
(Shannon's information), has been developed in terms of what could
be called a scientific theory. Among these meaning (probabilistic,
algorithmic, semantic, functional, etc) almost never appears the
physical aspects I will try to develop here: information as the
physical capacity of the carrier systems of modifying the physical
state of the receiver systems (what Küppers called the pragmatic
aspect of information \cite{Kuppers1990}).

As far as one can deduce from recent revisions of related subjects
\cite{{Eigen1992}, {Fox1988}, {Kauffman1993}, {Orgel1994},
{Stonier1990}, {Wicken1987}, {Yockey1992}} there is no formal
study of this undeniable physical capacity of the information
carrier systems. Although some ideas more or less related to
physical information can be found in the appropriate literature.
Among other examples, I will point out the following:

\begin{enumerate}

\item Kant's assumption that there exist in living organisms a
forming force with organizational capacity, which self-propagates
and cannot be explained in terms of simple mechanisms
\cite{Kant1984}.

\item Wolpert's positional information, referring to the spatial
position of the system's components, which is used in determining
the pattern of cellular differentiation \cite{Wolpert1978}.

\item Configurational information related to the configurational
microstates of the system \cite{Wicken1980}.

\item Margalef's suggestion of linking information to other
physical properties of the Universe, being a property of whatever
thing formed by different particles \cite{Margalef1980}.

\item Finally, Stonier work \cite{Stonier1990}, where information
is proposed as another essential constituent of the Universe
together with matter and energy.

\end{enumerate}

\noindent Stonier defines information as the capacity to organize
a system or to maintain it in an organized state. But he does not
take into account the arbitrariness of the established codes
between the carriers and the receivers, which is proposed here as
the essential feature of information. In addition, Stonier uses
the term organization as equivalent to order, which is very
debatable \cite{{Wicken1987}, {Wicken1989}}.

The discussion that follows will begin by analyzing the
inappropriateness of Shannon information and algorithmic
information for the study of the physical effects of the
expression of information. Immediately after it will be
demonstrated that it is possible to consider certain aspects of
the physical nature of the carrier systems in order to define a
new concept of (physical) information based on the capacity of
modifying the physical state of the receiver systems. Once defined
this new concept of physical information, the physico-mathematical
formalism will be applied to it, deriving as consequence the basic
properties of informed systems. These properties are immediately
recognized as the most characteristic attributes of living
systems. In this way, physical information introduces in Physics a
concept which is so strange to it as usual in biology: the concept
of function \cite{Wicken1987}. Physical information reveals itself
as the nexus of union between Physics and Functional Biology,
which, until now, maintain a certain mutual isolation
\cite{Wicken1985}. The discussion will finish by proposing a
double measure, physical and functional, of information. The
latter with an undoubtable evolutionary interest.

\section{Shannon Information}

\noindent Shannon information theory was developed by Hartley,
Shannon and Weaver in order to analyze the efficience of the
transmission of electronic signals \cite{Shannon1949}. Shannon
information measures the rarity of the messages formed by a
certain number of symbols, each one with an occurrence probability
of $p_i$. According to Shannon-Weaver's expression, the quantity
of information per symbol $(h)$ is given by:

\begin{equation}\label{eqn:Shannon expression}
h = -\sum_{i=1}^k p_i lg_2(p_i) \text{ bits}
\end{equation}

\noindent \\Once the quantity $h$ has been calculated, we can
immediately calculate the amount of information $(H)$ of any
message formed by $n$ of such symbols:

\begin{equation}\label{eqn:information n symbols}
H = nh
\end{equation}

\noindent \\Shannon information has been used with many different
purposes in biology. For example, it has been applied to analyze
nucleotide sequences \cite{{Brooks1988}, {Gatlin1972}}, biological
adaptation \cite{Conrad1983}, biological diversity
\cite{{Margalef1981}, {Margalef1983}}, ecological networks
\cite{{Ulanowicz1986}, {Wagensberg1990}}, quantitative ecology
\cite{Atlan1985}, even to establish the fundamentals of biological
evolution \cite{Brooks1988b}.

The biological success of Shannon information is probably due to
the mathematical analogy between expression (\ref{eqn:Shannon
expression}) and Boltzmann-Gibbs' expression for statistical
entropy. For certain authors, this mathematical analogy makes the
consideration of information as a form of entropy (or viceversa)
legitimate, and consequently, also legitimates the application of
entropy laws to information, specifically the second law of
thermodynamics. The validity of this supposed analogy between
information and entropy, and therefore that of its corresponding
biological applications, has been seriously contested
\cite{{Barry1986}, {Depew1988}, {Hariri1990}, {Olmsted1988},
{Shannon1949}, {Stuart1985}, {Wicken1987},{Wicken1987b}}.

What is not debatable is the lack of semantic value of Shannon
information. In effect, from equation (\ref{eqn:information n
symbols}) it immediately follows that all messages of equal length
have the same quantity of information, i.e. the same number of
bits. Shannon information has not physical significance either,
since different messages with the same Shannon information will
produce, in general, different effects in the receiver. Or in
other words, Shannon information does not distinguishes among
messages that produce different physical effects.

Let us consider a biological example. The gene that codes the
cytochrome $c$ in \emph{Rhodopseudomonas palustris} has 342
nucleotides. Therefore there could exist $4^{432}$ different genes
with the same Shannon information as it. Most of these genes would
produce proteins with different functional abilities from that of
the cytochrome $c$. Moreover, there could also exist a huge number
of genes with different Shannon information whose products would
be functionally equivalent to the considered cytochrome $c$. For
example, the gene that codes the cytochrome $c$ in
\emph{Pseudomonas mendocina} has 516 bits while that of
\emph{Rhodopseudomonas palustris} has 684. In consequence, from
the point of view of the functional ability of proteins the number
of bits of their coding genes is absolutely irrelevant.

Similar conclusions may be obtained from algorithmic information,
for which the information content of a message $m$ is equalled to
the number of bits of the shortest universal computer programme
that, when executed, yields as output the message $m$ in question
\cite{Zurek1989}. In this case, the lack of semantic and
functional value can be immediately deduced from the enormous
number of programs withe the same number of bits, but producing
different outputs.

Certain authors consider that although Shannon information
measurement is not related to meanings, it is sufficient to assume
that meanings exist, even though they are inaccessible to us as
external observers \cite{Atlan1985}. However, even when the
meaning is inaccessible to us, the physical changes produced in
the receiver as consequence of the expression of information can
be observed and also measured. The problem is that, in spite of
certain attempts \cite{Stonier1990}, we do not yet have a theory
of physical information. We do not even have a formal definition
of physical information.

Thus, we should speculate how this situation affects our
understanding of life. Since living beings are the result of the
physical changes produced through the expression of the
appropriate information on the appropriate receiver, can we expect
to understand their physical nature without a theory of physical
information? Can we expect to explain life leaving aside one of
its most singular attributes. It does not seem probable. On the
other hand, as we will immediately see, the consideration of the
physical aspects of information does not pose serious
difficulties.

\section{Physical Information}

\noindent As is well known, the information of a structural gene
is expressed by means of the cellular apparatus of translation,
provoking a physical change in its cytoplasmic receiver (basically
by converting a group of free amino-acids in a peptidical chain).
I will use this physical capacity of genes as a reference in order
to construct the definition of physical information. For this, it
seems convenient to recall the following well established facts:

\begin{enumerate}

\item The physical changes produced in the cytoplasm by the
expression of genes are arbitrary, i.e. they cannot be directly
deduced from physical laws, but from a code established, at least
partially, in an arbitrary way.

\item The process of translation is always the same, independently
of the translated sequence.

\item A given sequence of nucleotides always produces the same
change in the cytoplasmic receiver.

\item By simply changing the nucleotide sequence, different
changes in the cytoplasm will be obtained.

\item Genes are expressed again and again, while their nucleotide
sequences do not undergo any modification due to the expression
process.

\end{enumerate}

It is undeniable that genes have the strange capacity of changing
the physical state of their cytoplasmic receivers in many
different ways, although by the same basic process. Ii is
obviously about a physical capacity, since the changes produced by
them are of this nature. We also know that this capacity is base
on a configurational attribute of genes: their sequences of
nucleotides.

From a physical point of view, attention should be paid to the
arbitrariness of such changes, i.e. to the fact that the changes
cannot be deduced from physical laws, they cannot be predicted
without the knowledge of the genetic code. This is a truly
outstanding fact that, in my opinion, states the essential
difference between carrier and non-carrier information systems. In
effect, whilst it is theoretically possible to express the
behaviour of the latter in terms of mathematical equations deduced
from the physical laws driving their evolution, it is not possible
to do the same with the former: in this case we would also need to
know the arbitrary code that determines the changes in the
receiver.

This ability of genes, as strange as it might be, is real (genes
do exist). Undoubtedly, it deserves to be somehow termed, and it
seems adequate to do so by using the expression \emph{physical
information}. This is what I will propose in the following
discussion, where I will try to generalize and formalize these
suggestions. I will not use certain terms (like order,
organization, complexity) that may appear to be appropriate in the
subsequent discussion for two reasons. The first because they are
not necessary to the very discussion. The second because, although
they are usual terms, they are used with different meanings
\cite{Wicken1987}. Consequently, I would have to justify my own
selection of meanings, adding unnecessary debates. On the
contrary, the terms "system" and "configuration" will be used with
the maximum generality. As is usual in thermodynamics, I will use
the term \emph{system} to refer to \emph{any arbitrarily defined
part of the universe}, and the term \emph{configuration} in order
to refer to the \emph{spatial disposition of the arbitrarily
defined parts of a system}. I will also use the term
multi-configurational to refer to systems with the capacity of
exhibiting different configurations with the same components. We
are now in the appropriate situation to propose the following:

\begin{definition}\label{def:physical information}
Physical information is the capacity of producing arbitrary
changes.
\end{definition}

\noindent The systems that manifest this capacity will be termed
\emph{carriers}, while \emph{receivers} will be those that undergo
the changes. Although it it possible that both systems might be
independent from one another (although with a communication
channel between them), here I will deal only with carriers and
receivers integrated in only one system that I will term
\emph{informed system}. The decision is not capricious, not only
because this is the case for living organisms, but also for
reasons of feasibility \cite{Weber1989} and of stability (see
below the origin of information). I will use the term
\emph{expression} to refer to the set of interactions between the
carrier and the receiver of which results the change of state of
the latter. Finally, I will term \emph{apparatus of translation}
to the receiver components that carry out these interactions. From
now on, and whenever there is no confusion, I will use the term
\emph{information} to refer to the physical information that a I
have just defined.

The above definition states arbitrariness as the most outstanding
characteristic of the changes produced by information. This is
quite clear in human language, where the lack of physico-chemical
relations between a word and its designed object is so evident
\cite{Saussure1973}. The same applies to genetic language. The
affinity preferences between codons and their corresponding amino
acid found experimentally are not sufficient as to explain the
establishment of the genetic code (\cite{Aguilera1993},
\cite{Crick1968} and references therein, \cite{Kauffman1993},
\cite{Moras1990} and references therein). Both the existence of
homonymous codes \cite{{Ayala1984}, {Berry1991}, {Kawaguchi1989},
{Rennie1993}} and the distance separating the anticodon site from
the site where aminoacids bind tRNA, support this experimental
conclusion \cite{Kauffman1993}. Although some initial prebiotic
interactions, probably hydrophobic (\cite{Aguilera1993} and
references therein), between groups of codons and groups of
aminoacids could exist, it seems clear that at least a certain
arbitrariness has remained frozen in the genetic code. On the
other hand, the following formal deduction of the physical
properties of informed systems is independent of the grade of
arbitrariness, provided that this grade may not be null.

\section{Physical Properties of Informed Systems}

\noindent As with all physical capacity, information must also be
based on some physical attribute of its carrier system. Since
arbitrary changes are determined by an arbitrary code rather than
by physical laws, an attribute capable of producing arbitrary
changes must comply with the following requirements:

\begin{enumerate}

\item It must be capable of producing different changes in the
receiver by the same type of process. Otherwise the changes would
not be arbitrary but determined by physical laws.

\item It must be multiconfigurational of wide range, otherwise it
could not be the physical support of the enormous information
diversity.

\item Its different configurational states must have the ability
of producing different physical changes in the receiver, while the
same configurational state will always produce the same change.

\item Arbitrariness requires that the current correspondence
between the carriers' configurational states and the receiver
physical changes be not the only possible.

\item It must be a stable attribute with respect to the expression
processes. If not, information would be destroyed and would have
to be created continually.

\end{enumerate}

\noindent These conditions imply the following

\begin{proposition}
Information carrier systems must be multiconfigurational and
stable in the face of the expression processes.
\end{proposition}

\noindent The carrier configurational attribute on which
information is based, would not be operative if its different
configurational states do not have the opportunity of being
originated approximately with the same probability. However, it is
not necessary that these configurational changes have to occur in
any special sequence nor in any special way. It is sufficient that
they are occasionally produced. It is enough therefore, that such
changes may be originated through sporadic and random interactions
between the carrier and its environments. Thus, we can state the
following:

\begin{proposition}
Information carrier systems must have access to their different
configurational states at least through random interactions with
their environment.
\end{proposition}

\noindent In natural systems, arbitrariness can only arise from
the fixation of randomly selected processes. In consequence, since
the very fixation is an irreversible process, it can only occur in
removed from thermodynamic equilibrium systems. Therefore,
informed systems must be produced and maintained away from
equilibrium. This circumstance requires the continuous supply of
energy, which in turn demands that informed systems have to be of
a non isolated nature, with the ability of maintaining energetic
interchanges with their environment. In addition to import energy
the maintenance of a non equilibrium steady state also requires
that the system export to its surroundings all the entropy that
its irreversible processes inevitably produce. Accordingly, we can
state the following:

\begin{proposition}\label{prp:non isolated}
Informed systems must be of a non isolated nature, able to
maintain their non equilibrium states by interchanging energy and
entropy with their environments.
\end{proposition}

\noindent This necessity of interchanging energy and entropy
involves the nature of the environment in the maintenance of
informed systems. As we will see next, some of the most
interesting properties of informed systems are derived from this
circumstance. In order to deduce them, the following discussion
about the stability of removed from equilibrium systems is
necessary.

Consider any system with the ability stated in Proposition
\ref{prp:non isolated}. The phenomenological relationships ($F_i$)
linking the flows ($J_i$) of matter and energy with their
generalized driving forces ($X_i$):

\begin{equation}
J_i = F_i(X_1, X_2, \dots X_n)
\end{equation}

\noindent \\are generally non lineal, and therefore  very
sensitive to small variations in the intensity of the forces. On
the other hand, the stability conditions for non equilibrium
steady states are given by:

\begin{align}
\delta^2S &< 0 \label{eqn:stability 1}\\
\frac{d(\delta^2S)}{dt} & \geq 0 \label{eqn:stability 2}
\end{align}

\noindent \\where $\delta^2S$ denotes the excess of entropy
\cite{Glansdorff1971}. Internal fluctuations around a non
equilibrium steady state arise in the system with a probability:

\begin{equation}
P = e^{\delta^2S/2K}
\end{equation}

\noindent \\where $K$ is Boltzmann's constant. While conditions
\ref{eqn:stability 1} and \ref{eqn:stability 2} hold, all small
fluctuations in the intensity of the forces are damped down and
the system recovers its steady state. Otherwise, any small
fluctuation can be amplified and the system may spontaneously
evolve to another macroscopic state, including the equilibrium
one.

Equation \ref{eqn:stability 2} represents the dynamic factor's
contribution to stability, such as the rate at which matter and
energy are supplied to the system. Because of environmental
fluctuations, variations of these energetic material supplies are
unavoidable. Some of these variations may be the cause for which
equation \ref{eqn:stability 2} no longer holds, and consequently
the cause for which the system becomes unstable, sensitive to
small fluctuations. Let $p_i(t)$ be the probability for an
environmental fluctuation to occur during a time $t$ causing a
change in the flow $J_i$ that the system decays to equilibrium.
The fluctuating behaviour of nature allow us to state that:

\begin{align}
p_i(t) > 0 \label{eqn:pi > 0}\\
\frac{d(p_i(t))}{dt} > 0 \label{eqn:d(pi(t))/dt > 0}
\end{align}

\noindent \\Hence, the self-maintaining probability $P(t)$ of an
informed system sustained at the expense of $n$ of these matter
and energy flows may be written as:

\begin{equation}
P(t) = \sum_{i=1}^{n}\left[1 - p_i(t) \right]
\end{equation}

\noindent \\and according to (\ref{eqn:d(pi(t))/dt > 0}):

\begin{equation}
\frac{dP(t)}{dt} = - \sum_{i=1}^n \frac{dp_i(t)}{dt}\sum_{j=1}^n(1
- p_{j, \ j \neq i}(t)) < 0
\end{equation}

\noindent \\i.e. $P(t)$ decreases with time, and we cannot ensure
the permanence of any informed system over sufficiently long
periods of time. Moreover, if $dp_i(t)/dt$ are not decreasing
functions -and there is no reason to assume that they are- then it
can be easily shown that $d^2P(t)/dt^2 > 0$, which means that
$P(t)$ decreases exponentially with time. In conclusion, we cannot
ensure the existence of natural informed systems.

The only way to solve this situation is to assume that informed
systems have self-replicating ability. This is not a strange
capacity: living beings are self-replicating systems, and they are
not the only possible ones \cite{Rebek1994}. Let us assume the
existence of $x$ self-replicating systems at a given time $t$, we
can write the following demographic equation:

\begin{equation}
x_{n+1} = r p x_n
\end{equation}

\noindent \\where $x_{n+1}$ and $x_{n}$ denote the number of
systems of two successive generations; $r$ the replication rate
(the number of copies per system); and $p$ the probability that
the system achieve its replication. The maintenance of the number
of systems of the group requires that:

\begin{equation}
x_{n+1} = x_n
\end{equation}

\noindent \\and therefore that:

\begin{equation}
rp = 1
\end{equation}

\noindent whence:

\begin{equation}
r > 1
\end{equation}

\noindent \\since $p < 1$. Thus, we can state the following:

\begin{proposition}
It is only possible to ensure the existence of informed systems of
a self-replicating nature with a replication rate greater than 1.
\end{proposition}

\noindent Evidently, self-replicating requires the use of
environmental materials for the production of copies. That is to
say, self-replicating systems must be open systems, being able to
maintain exchanges not only of energy but also of matter wit their
surroundings. From now on, I will deal only with these
self-replicating informed systems.

Let us term generation renewal time ($t_g$) as the time lapsed
between two successive replications of the components of a group
of informed systems. Since environmental fluctuations occur with a
high irregular frequency, not all the interval $t_g$ will be
equally favorable for replication. In other words, the replication
probability depends on the considered period of time. Let

\begin{align}
x_{n+1} &= rpx_n \\
rp &= 1
\end{align}

\noindent \\be the demographic equations of a group of informed
systems, $p$ being referred to any period of time $t_g$. At less
favorable periods, the replication probability $p_L$ will be
smaller than $p$ and, consequently, the number of systems will
decrease:

\begin{equation}
x_{n+1} = rp_Lx_n < x_n
\end{equation}

\noindent since

\begin{equation}
rp_L < rp = 1
\end{equation}

\noindent \\At more favorable periods, we will have the opposite
situation ($p_M > p$) and the number of systems increases:

\begin{equation}
x_{n+1} = rp_M x_n > x_n
\end{equation}

\noindent since now

\begin{equation}
rp_M > rp = 1
\end{equation}

\noindent \\If the recovery is not sufficient the group may be
extinguished after a seres of unfavorable periods. Therefore the
only groups which ensure their maintenance are those for which $rp
> 1$ most of the time. But in this type of system a new class of
interaction appears. In fact, let $N_0$ be the initial number of
informed systems of one of these groups. At the end of $k$
favorable consecutive generations the number of elements of the
group will be:

\begin{equation}
N_k = (rp)^kN_0; \ rp > 1
\end{equation}

\noindent \\The above exponential growth in the number of systems
would also demand an exponential growth of the environmental
material and energetic resources, which is impossible. Thus, the
limitation of environmental resources imposes restrictions in the
number of systems of the group, and therefore, the appearance of
demographic competitive interactions: in many generations more
copies are formed than can be sustained by the environment.
Consequently, we come to:

\begin{proposition}
Informed systems are liable to undergo extinctions and demographic
competitive interactions.
\end{proposition}

\noindent Let us consider now any group of informed systems. In
general, the carriers of different systems of the group will
exhibit different configurational states that, we can assume, have
been produced by simple random interactions with their
environment. Naturally, the only possible configurational states
of the carriers will be those which produce changes in the
receivers compatible with their necessity for maintenance and
replication. But not all of them will be equally efficient in
relation to such activities. Let us assume, for the sake of
simplicity, that our group is formed by two types of systems, $A$
and $B$, so that their respective replication probabilities
verify:

\begin{equation}\label{eqn:pA < pB}
p_B < p_A
\end{equation}

\noindent \\Let $A_k$ and $A_{k+1}$ be the number of $A$-systems
in two successive generations. Its relative variation rate ($d_A$)
can be written as:

\begin{equation}
d_A = \frac{A_{k+1}-A_k}{A_k}; \ A_k > 0
\end{equation}

\noindent \\and taking into account that:

\begin{equation}
A_{k+1} = rp_AA_k
\end{equation}

\noindent \\we can write:

\begin{align}
d_A &= \frac{rp_AA_k - A_k}{A_k}\\
&= rp_A -1
\end{align}

\noindent \\Similarly, for $B$-systems we come to:

\begin{equation}
d_B = rp_B - 1
\end{equation}

\noindent \\Finally, from (\ref{eqn:pA < pB}) it follows that:

\begin{equation}
d_B = rp_B - 1 < rp_A - 1 = d_A
\end{equation}

\noindent \\which means the relative increase of $A$-systems with
respect to $B$-systems. And consequently, the increase of the
average replication probability of the whole group. Note that this
increase does not depend on the demographic state of the group,
which may be stationary or variable. Therefore, we can state the
following:

\begin{proposition}\label{prp:increase replication}
Informed systems continually increase their replication
probability, until reaching the maximum value compatible with
their environmental conditions.
\end{proposition}

\noindent The above proposition states the existence of a
directional, evolutionary process in informed systems. In reality,
this proposition is a physical and formalized version of Darwin's
Principle of Natural Selection.

Finally, we will consider the anisotropy of the natural
environment. That is to say, the qualitative and quantitative
differences in its properties. These differences allow us to
consider the natural environment as formed by a set of mosaics
differentiated by some properties which are significant from the
point of view of the maintenance and replication of informed
systems. Now, all we have to do is to apply Proposition
\ref{prp:increase replication} to the systems of each one of these
mosaics, thus deducing that he changes which contribute to
increase the replication probability of the systems under the
particular conditions of each mosaic are to be propagated. We thus
have the following:

\begin{proposition}
The environmental anisotropy produces the diversification of
informed systems.
\end{proposition}

It seems appropriate now to make a brief r\'esum\'e and an
evaluation of the results obtained until this point. We have
proposed a definition of physical information and we have applied
to it the physco-mathematical formalism. By doing so, we have
deduced the most basic physical properties of informed systems,
that can be resumed in the following way. Informed systems:

\begin{enumerate}

\item have multiconfigurational carriers, which undergo changes in
their configurational states.

\item are open systems, capable of interchanging matter, energy
and entropy with their environment.

\item are self-replicating systems, with a replication rate
greater than one.

\item are exposed to undergo extinctions and competitive
interactions.

\item suffers evolutionary processes.

\item suffers diversification processes.

\end{enumerate}

\begin{figure}[htb!]
\begin{center}
\includegraphics[scale=0.8]{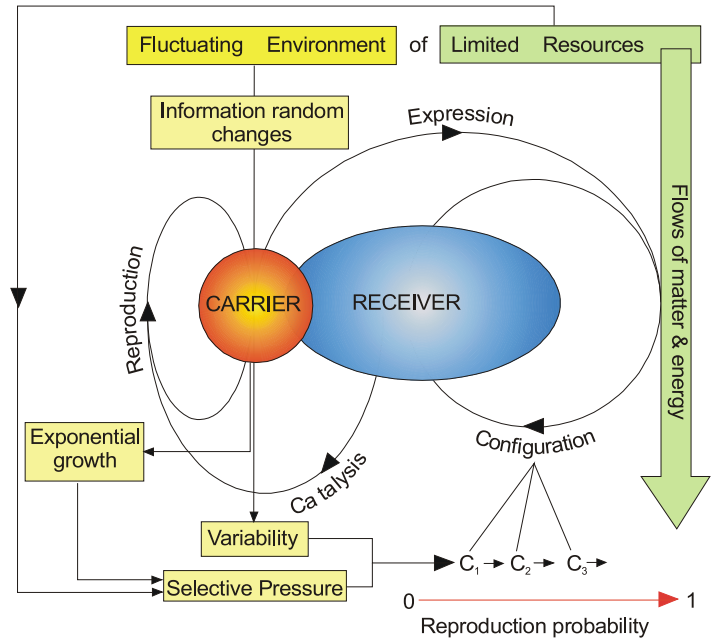}
\caption{Living beings as informed system: operating and
evolution}\label{fig:living beings}
\end{center}
\end{figure}

\noindent It is obvious to recognize in these properties the most
significant attributes of living beings, which are the only
natural systems we know of (Figure \ref{fig:living beings}).
Although no new property of living beings has been discovered
here, the method used to obtain them is a significant novelty: the
physico-mathematical formalism applied to a definition of
information that consider it as a physical entity rather than a
theoretical one.

In spite of the conflict that may result from proposing the
existence of a new physical entity, in the case of information the
only thing is to do it. Living organisms exist and, in effect,
their genes show the capacity that we have defined as physical
information. In addition, the proposed definition is very close to
that of energy (the capacity of producing changes), the only
significant difference being the arbitrariness of the changes
produced by information in the face of the physical determinism of
the energetic changes. Information could be really considered as a
sort of energy, which arises in system removed from thermodynamic
equilibrium, as a consequence of the fixation of certain randomly
selected processes. To compensate, this new physical concept will
allow us that living organisms no longer be these historical
objects so strange to physics \cite{{Bohm1976}, {Nicolis1989},
{Prigogine1983}, {Prigogine1990}, {Prigogine1990b}}. It will also
allow the development of the physico-mathematical formalism
necessary in my opinion, to end once and for ll the controversy
permanently rooted in the very foundations of evolutionary biology
\cite{{Dover1992}, {Hellier1991}, {Maddox1991}, {Maynard1976},
{Weale1991}}. In addition to the qualitative explanations, we will
immediately see that it is also possible to obtain quantitative
results related to the physical and functional effects of
information.

\section{Information and Functional Activity}

\noindent The concept of function is related to the activity of
certain systems such artificial machines or living beings. It can
be considered as the objective or purpose that the system has to
attain. Although it can also be considered as the activity
developed in order to attain the objective in question (functional
activity). Here, I will use the term with the maximum generality:
a \emph{function} will be \emph{any arbitrarily defined purpose}.
In natural nonliving objects, there is no sense is speculating
about their objective or purposes, since they are nothing more
than the products of the evolution of the universe under the
driving force of physical laws. For this reason, the concept of
function does not belong to the lexicon of physics
\cite{Wicken1987}. As we will see, in the case of an informed
system there are objectives or functions for which it makes sense
to speculate.

In effect, we already know that informed systems have arbitrary
characteristics (which are produced by the joint action of the
random modifications of the carriers and the arbitrary
interactions between the carriers and the receivers). Although
these characteristics are not deducible from physical laws, they
are the same laws which determine their permanence or extinction.
Because they are those which govern the behavior of the natural
environment where informed systems try self-maintaining and
self-replicating, and those which impose restriction to the very
physico-chemical processes involved in such activities of
maintenance and replication. One can say that physical laws put to
test the arbitrary attributes of informed systems. It makes sense,
therefore, to ask whether a given informed system will reach the
objective of its own replication. The answer is not obvious, and
the lack of obviousness justifies the physical consideration of
the concept of function and that of functional activity. So,
informed systems introduce in Physics these concepts so typically
biological.

Naturally, each function will be attained from a certain state of
the system, which in turn will be the result of the expression of
its information. In this way it will be possible to evaluate
information from the point of view of the functional ability of
the system. From a practical point of view, it may be interesting
to decompose a complex function (as presumably is
self-replicating) in an ordered series of subfunctions, each one
of them attained from a certain state of the system, which in turn
results from the expression of a certain "fragment" of
information. For this to be possible, the system has to be able
-as occurs in living beings- to express an ordered series of
fragments of information, which does not add any new difficulty of
its functioning. It is sufficient to consider -again as in living
organisms- the appropriate interactions between certain components
of the receiver and certain special fragments of information
(metainformation) \cite{{Beato1991}, {Manak1993}, {Phillips1993}},
which, finally, will determine the expression of the adequate
fragment of information.

\section{Physical and Functional Measurements of
Information}

\noindent As I have repeatedly said, the expression of information
modifies the state of its receivers systems. Next, I will propose
a way of measuring the intensity of these modifications.
Statistical thermodynamics provides us with the appropriate
grounds for such an objective. In effect, as is well known, a
change in the state of a system involves a change in the number of
its compatible microstates. Thus we can estimate the physical
effects of information in terms of the variation in the number of
such microstates suffered by the system because of the expression
of its information.

Consider two macrostates of an informed system. The first that of
thermodynamic equilibrium, the second the resulting from the
expression of its information. Let $W_e$ and $W_i$ be the number
of microstates which are respectively compatible with such
macrostates. I will define the measurement of physical
information, $\Omega$, as:

\begin{equation}\label{eqn:physical information}
\Omega = \ln\left(\frac{W_e}{W_i}\right)
\end{equation}

\noindent \\The quantity $\Omega$ is a dimensionless measurement
that represents the decreasing rate in the number of microstates
caused by the expression of information. In other words, it
represents the degree of removal from thermodynamic equilibrium
caused by information (it could be said to be the degree of
"informization"). It is immediate to express $\Omega$ in terms of
entropy variations. Indeed, according to Boltzmann's equation we
can write:

\begin{align}
S_e = K \ln W_e \label{eqn:Se}\\
S_i = K \ln W_i \label{eqn:Si}
\end{align}

\noindent \\where $S_e$ and $S_i$ denote the system entropy at the
considered conditions, and $K$ is Boltzmann's constant. From
(\ref{eqn:Se}) and (\ref{eqn:Si}) we immediately come to:

\begin{equation}
\Delta S = S_i - S_e = K \ln \left(\frac{W_i}{W_e} \right)
\end{equation}

\noindent hence:

\begin{equation}\label{eqn:We/Wi}
\frac{W_e}{W_i} = e^{(-\Delta S / K)} = P^{-1}
\end{equation}

\noindent \\where $P$ is Einstein's probability, which is the
probability for a fluctuation to arise in a system at equilibrium,
originating the change $\Delta S$ in the entropy of the system.
According to (\ref{eqn:We/Wi}) we can write (\ref{eqn:physical
information}) as:

\begin{equation}\label{eqn:Omega 1}
\Omega = \frac{- \Delta S}{K}
\end{equation}

\noindent or:

\begin{equation}\label{eqn:Omega 2}
\Omega = - \ln P
\end{equation}

\noindent \\since $\Omega$ is a dimensionless quantity, we can
define ay arbitrary unit of measurement. For instance, we can
define the physical information unit (pit) as the quantity of
information for which $W_i = 1/2 W_e$. That is:

\begin{equation}
\text{1 pit } = \ln \left( \frac{W_e}{(1/2)W_e} \right) = \ln 2
\end{equation}

\noindent \\Therefore (\ref{eqn:Omega 1}) and (\ref{eqn:Omega 2})
can be rewritten as:

\begin{equation}
\Omega = - \frac{\Delta S}{K} \ln 2 \text{ pits}
\end{equation}

\noindent and:

\begin{equation}
\Omega = - \lg_2 P \text{ pits}
\end{equation}

\noindent \\Note that the physical meaning of $\Omega$ is not
debatable: it has been defined explicitly in terms of entropy
variations.

As is to be expected, the values of $P$ in living organisms are
extremely small. For instance, in the case of \emph{Escherichia
coli} the probability of a living microstate is in the order of
$10^{-10^{11}} = 10 ^{-100000000000}$ \cite{Morowitz1968}. The
changes in entropy $\Delta S$ can be estimated, at least for
single unicellular organisms, by micro-calorimetry
\cite{Lurie1979} (an estimate of $-3.35 \times 10^{-12} JK^{-1}$
in the variation of entropy during the process of the formation of
a cell was found empirically by Luri\'e and Wagensberg;
accordingly, the quantity $\Omega$ involved in the process would
be of $3.5 \times 10^{11}$ pits).

In addition to measuring the global effects of information, we can
measure the partial effects of the expression of a given fragment
of information. We can use the same definition (\ref{eqn:physical
information}):

\begin{equation}
\Omega_{12} = \ln \left( \frac{W_1}{W_2} \right)
\end{equation}

\noindent \\$W_1$ and $W_2$ now denoting the number of microstates
of the receiver before and after, respectively, the expression of
the considered information.

The measure of physical information we have just defined, although
interesting from the theoretical point of view, has no functional
value. In fact, due to the arbitrariness of the changes that
information produces in its receiver, many of them can produce the
same variations in the number of microstates (the same $\Omega$)
but very different functional abilities in the receiver. On the
other hand, it is easy to establish a functional measurement of
information, which could also be interesting form the semantic
point of view. For this we need to know the function for which
information is evaluated. In fact, let $u$ be any function defined
for an informed system, and let $p$ be the probability of
achieving this function once expressed the information $I$ on a
certain receiver state. I will define the functional measure,
$\phi$ of $I$ as:

\begin{equation}
\phi = f(p)
\end{equation}

\noindent \\where $f$ is a type $C^2$ (continuous function whose
first and second derivatives are also continuous) real function
defined on $[0, 1]$ such as:

\begin{align}
&f(p) = 0 \Leftrightarrow p = 0 \label{eqn:functional 1} \\
&f'(p) > 0, \ \forall p \in [0, 1] \label{eqn:functional 2}
\end{align}

\noindent \\which states that $\phi$ increases strictly as the
probability $p$ of attaining the function in question increases.
Due to the extremely complex non-lineal dynamics of nature, it
seems reasonable to assign an infinite value to the measurement
$\phi$ which corresponds to the information capable of assuring
the attainment of the function $u$ in any possible environmental
conditions ($p = 1$). Thus, we might also require:

\begin{align}
&f(1) = \infty \label{eqn:functional 3}\\
&f"(p) > 0 \label{eqn:functional 4}
\end{align}

\noindent \\Any mathematical function satisfying
(\ref{eqn:functional 1}) and (\ref{eqn:functional 2}) could be
used to measure functional information. I propose the following:

\begin{equation}
\phi = \frac{p}{1-p}
\end{equation}

\noindent \\which also satisfies (\ref{eqn:functional 3}) and
(\ref{eqn:functional 4}). The quantity $\phi$ is a dimensionless
measurement too, thus we can define any arbitrary unit for it. For
instance we can define the functional information unit (fit) as
the quantity $\phi$ for which $p = 1/2$. That is to say:

\begin{equation}
\text{1 fit }= f(1/2) = \frac{1/2}{1 - 1/2} = 1
\end{equation}

\noindent therefore:

\begin{equation}
\phi = \frac{p}{1-p} \text{ fits}
\end{equation}

\noindent \\From the empirical point of view, the functional
measurement of information results are much more interesting than
the physical one. For living beings, beside the basic function of
self-reproducing, many other biological functions can be used to
evaluate their genetic information in the quantitative,
probabilistic terms we have just defined.

\section{The Origin of Information}

\noindent Although it is not an objective of this paper, it seems
appropriate to make some considerations on the origin of
information. As is to be expected, this question is equivalent to
that of the origin of life from the perspective of physical
information. I will simply show that this new perspective does not
add any new problem.

Let us begin by asking ourselves about the possible nature of the
carrier and the receiver systems. The same quantum-mechanical
reasons from which Schr\"odinger derives the molecular nature of
its "aperiodic crystals" \cite{Schrodinger1967}, will serve us to
deduce the molecular nature of the carrier systems. In effect,
only molecules could reach the size, variety and stability
required by a multiconfigurational system. Carrier molecules have
to be able to acquire different equiprobable configurations, and
hence they have to be formed by at lest two different components
with the appropriate binding ability so as to constitute the
carrier in question. These components cannot be individual atoms
simply because different atoms will bind to one another with
different bonds, which is incompatible with the required
equiprobability of the resulting arrangements. The components must
be molecules able to form at least two equivalent chemical bonds
allowing equiprobable configurations. That is to say, they have to
be copolymers made up of two or more different types of monomers,
the sequence of such monomers being the multiconfigurational
attribute upon which the carrier capacity we have termed physical
information is based.

Consequently, the receivers must also consist of molecules, which
should be sensitive to the different carrier sequences, in the
sense that their different states must be arbitrarily determined
by the sequence of carrier monomers, rather than by any type of
spatial complementarity or physico-chemical affinity between them.
One way for this to be possible is that the receiver be also
formed by copolymers. In these circumstances the process of
expression would consist of translating the sequence of carrier
monomers into a sequence of receivers monomers through an
arbitrary code and an appropriate molecular apparatus of
translation. Thus, the first problem we have to deal with is about
the origin of such monomers and polymers. This is a classical
problem in protobiogenesis on which important researches exist,
whose conclusions can be reviewed in any of the numerous text
devoted to the subject (for instance \cite{Chyba1992},
\cite{Day1986}, \cite{Fox1988}, \cite{Garcia1985}, \cite{Fox1972},
\cite{Horgan1991}, \cite{Orgel1975}, \cite{Orgel1994}). In
summarizing, we can recall that such molecules could have been
formed, with more or less probability, on the early Earth.

Once the polymers were available, the appropriate interactions
among them had to settle down so as to become living (informed)
systems. We can separate these interactions into two different
groups: catalytic interactions driving the process of carrier
replication; and informed interactions consisting in the
translation of the carrier monomer sequence into the receiver one.
Both types of interactions are being studied intensely at present.
There are several models for the first type of interaction, among
them the model of hypercycles \cite{{Eigen1979}, {Eigen1992}} and
the model of autocatalytic sets of catalytic polymers
\cite{Kauffman1993}. The main problem of the second type of
interaction is the origin of the genetic code, for which there are
some interesting suggestions, such as those of Crick
\cite{Crick1976} and Bedian \cite{Bedian1982}. Many of these
investigations reveal the astonishing emergent properties of
systems subjected to iterative processes, including the emergence
of stabilizing attractors in systems whose components suffer
iterative cycles of mutual interactions \cite{{Bak1991},
{Bak1989}, {Hofstadter1982}, {Kauffman1992}, {Leon1990},
{May1976}, {May1991}, {Sigmund1993}, {Stewart1989}, {Zeleny1977}}.
All these suggest an origin of living (informed) system more
expected \cite{Kauffman1993} than miraculous \cite{Monod1970}.
Thus we can conclude that the origin of life, and consequently
that of physical information, although still unsolved, is a
scientific question rather than a metaphysical one.

\section{Conclusions}

\noindent The consideration of information as the capacity of
producing arbitrary changes is not only necessary (genes do exist
and they show just that ability) but also productive: it allows to
deduce in a formal way the most significant attributes of living
beings. Thus, information introduces in Physics a property so
strange to the non-living world as genuine in living ones:
arbitrariness. By the simple consideration of this property, life
can be explained in physical terms. As we have just seen, present
day researches on the origin and evolution of life suggest more
and more that living (informed) systems, rather than a miracle of
chance, are expected consequences of physical laws. Life seems to
be nothing more than random and arbitrariness under the effects of
physical laws.

We have also seen how the physical effects of information can be
doubly evaluated. On the one hand it is possible, at least
theoretically, to estimate the intensity of the changes suffered
by the receiver in terms of entropy variations. On the other hand,
it is also possible to evaluate in probabilistic terms its
functional capacity. That is to say, its ability to reach a given
objective. The proposed functional measurement of information is
of practical use and may be of evolutionary interest, since it
allows us to evaluate quantitatively genetic information.

\noindent \\ \textbf{Acknowledgement }

\noindent I would like to thank Dr. Martínez García-Gil for his
stimulating discussion.


\providecommand{\bysame}{\leavevmode\hbox
to3em{\hrulefill}\thinspace}
\providecommand{\MR}{\relax\ifhmode\unskip\space\fi MR }
\providecommand{\MRhref}[2]{%
  \href{http://www.ams.org/mathscinet-getitem?mr=#1}{#2}
} \providecommand{\href}[2]{#2}

\end{document}